\documentclass[conference,a4paper]{IEEEtran}
\IEEEoverridecommandlockouts 
\usepackage{amsmath,graphicx}
\usepackage{cite}
\usepackage{balance}
\usepackage{multirow}
\usepackage{makecell}
\usepackage{pgfplots}
\usepackage{amssymb,amsfonts}
\usepackage{algorithmic}
\usepackage{xcolor}
\usepackage{boldline, soul}
\usepackage{booktabs}
\usepackage{indentfirst}
\usepackage{enumitem}
\usepackage{pifont}
\usepackage{pdfpages}
\usepackage{textcomp}
\usepackage[acronyms]{glossaries}
\usepackage{algorithm}
\usepackage{algorithmic}
\usepackage{caption}
\def\BibTeX{{\rm B\kern-.05em{\sc i\kern-.025em b}\kern-.08em
    T\kern-.1667em\lower.7ex\hbox{E}\kern-.125emX}}
    
    \setcellgapes{3pt}
\def\BibTeX{{\rm B\kern-.05em{\sc i\kern-.025em b}\kern-.08em
    T\kern-.1667em\lower.7ex\hbox{E}\kern-.125emX}}

\pgfplotsset{compat=1.13}

\newcolumntype{P}[1]{>{\centering\arraybackslash}p{#1}}
\newcolumntype{M}[1]{>{\centering\arraybackslash}m{#1}}
\newcolumntype{C}[1]{>{\raggedright\arraybackslash}m{#1}}

\bstctlcite{IEEEexample:BSTcontrol}

\makeglossaries
\newacronym{cnn}{CNN}{Convolutional Neural Network}
\newacronym{avc}{AVC}{Advanced Video Coding}
\newacronym{vvc}{VVC}{Versatile Video Coding}
\newacronym{hevc}{HEVC}{High Efficiency Video Coding}
\newacronym{dbf}{DBF}{De-Blocking filter}
\newacronym{qp}{QP}{Quantization Parameter}
\newacronym{sao}{SAO}{Sample Adaptive Offset}
\newacronym{alf}{ALF}{Adaptive Loop Filter}
\newacronym{sr}{SR}{Super Resolution}
\newacronym{pp}{PP}{Post Processing}
\newacronym{ilf}{ILF}{In-Loop Filtering}
\newacronym{qe}{QE}{Quality Enhancement}
\newacronym{gop}{GOP}{Group of Pictures}
\newacronym{evc}{EVC}{Essential Video Coding}
\newacronym{ctu}{CTU}{Coding Tree Unit}
\newacronym{ctb}{CTB}{Coding Tree Block}
\newacronym{mse}{MSE}{Mean Squared Error}
\newacronym{cu}{CU}{Coding Unit}
\newacronym{ipm}{IPM} {Intra Prediction Modes}
\newacronym{tu}{TU}{Transform Unit}
\newacronym{pu}{PU}{Prediction Unit}
\newacronym{msssim}{MS-SSIM}{Multi-Scale Structural SIMilarity}
\newacronym{pqf}{PQF}{Peak Quality Frame}
\newacronym{mc}{MC}{Motion Compensated}
\newacronym{gpu}{GPU}{Graphic Processing Unit}
\newacronym{ai}{AI}{All Intra}
\newacronym{ra}{RA}{Random Access}
\newacronym{ld}{LD}{Low Delay}
\newacronym{vr}{VR}{Virtual Reality}
\newacronym{dc}{DC}{Direct Current}
\newacronym{bd-br}{BD-BR}{Bj\o ntegaard Delta Bit Rate}
\newacronym{ct}{CT}{Coding Type} 
\newacronym{rt}{RT}{Run-Time}
\newacronym{cbr}{CBR}{Constant Bit-Rate}
\newacronym{mv}{MV}{Motion Vector}
\newacronym{psnr}{PSNR}{Peak Signal-to-Noise Ratio}
\newacronym{ctc}{CTC}{Common Test Condition} 
\newacronym{jvet}{JVET}{Joint Video Experts Team} 
\newacronym{ms}{MS}{Model Selection}
\newacronym{tum}{TUM}{Transform Units Mask}
\newacronym{cum}{CUM}{Coding Units Mask}
\newacronym{resm}{ResM}{Residual Mask}
\newacronym{predm}{PredM}{Prediction Mask}
\newacronym{mm}{MM}{Mean Mask}
\newacronym{ipmm}{IPMM}{Intra Prediction Mode Map}

\newacronym{dcad}{DCAD}{Deep CNN-based Auto Decoder}
\newacronym{IFCNN}{IFCNN}{In-Loop filter CNN }
\newacronym{STResNet}{STResNet}{Spatial-Temporal Residue Network}
\newacronym{DCAD}{DCAD}{Deep CNN-based Auto Encoder}
\newacronym{DSCNN}{DSCNN}{Decoder-side Scalable Convolutional Neural Network}
\newacronym{MMS-net}{MMS-net}{Multi-modal/Multi-scale Convolutional Neural Network}
\newacronym{VRCNN}{VRCNN}{Variable-filter-size Residue-learning CNN}
\newacronym{MSDD}{MSDD}{Multi-Scale deep decoder}
\newacronym{QECNN}{QECNN}{Quality Enhancement Convolutional Neural Network}
\newacronym{MFQE}{MFQE}{Multi-Frame Quality Enhancement }
\newacronym{CNNF}{CNNF}{CNN Filter}
\newacronym{RHCNN}{RHCNN}{Residual Highway CNN}
\newacronym{FECNN}{FECNN}{Frame Enhancement CNN}
\newacronym{Residual-VRN}{Residual-VRN}{Residual-based Video Restoration Network}
\newacronym{MGANet}{MGANet}{Multi-frame Guided Attention Network}
\newacronym{MLSDRN}{MLSDRN}{Multi-Channel Long-Short term Dependency Residual Network}
\newacronym{ADCNN}{ADCNN}{Attention based Dual-scale CNN}
\newacronym{MM-CU-PRN}{MM-CU-PRN}{Multi-scale Mean value of CU-Progressive Rethinking Network}
\newacronym{SDTS}{SDTS}{Spatial Details and Temporal Structure}
\newacronym{VRCNN-BN}{VRCNN-BN}{Variable Filter-size Residue-learning CNN with batch normalization layer}
\newacronym{MIF}{MIF}{Multi-frame In-loop Filter}
\newacronym{MRRN}{MRRN}{Multi-Reconstruction Recurrent Residual Network}
\newacronym{RRCNN}{RRCNN}{Recursive Residual Convolution Neural Network}
\newacronym{B-DRRN}{B-DRRN}{Deep Recursive Residual Network with Block information}
\newacronym{CPHER}{CPHER}{Coding Prior based High Efficiency Restoration}
\newacronym{MPRN}{MPRN}{Multi-level Progressive Refinement Network}
\newacronym{BDRRN}{BDRRN}{Block Information Constrained Deep Recursive Residual Network}
\newacronym{SEFCNN}{SEFCNN}{Squeeze-and-Excitation Filtering CNN}
\newacronym{GAN}{GAN}{Generative Adversarial Network}
\newacronym{MFRNet}{MFRNet}{Multi-level Feature review Residual dense Network}
\newacronym{IResLB}{IResLB}{Inception and Residual Learning based Block}

\begin{document}

\title{Model Selection CNN-based VVC Quality Enhancement}

\author{

\IEEEauthorblockN{Fatemeh Nasiri $^{\star\dagger+}$, Wassim Hamidouche$^{\star\dagger}$, Luce Morin$^{\dagger\star}$, Nicolas Dhollande$^{+}$, Gildas Cocherel$^{+}$}

\IEEEauthorblockA{$^{\star}$ IRT b$<>$com, 35510 Cesson-S\'{e}vign\'{e}, France,\\
		 $^{\dagger}$ Univ Rennes, INSA Rennes, CNRS, IETR - UMR 6164, 35000 Rennes, France \\
		 $^+$ AVIWEST, 35760, Saint-Gr\'{e}goire, France
		 } 
}

\maketitle

\begin{abstract}
Artifact removal and filtering methods are inevitable parts of video coding. On one hand, new codecs and compression standards come with advanced in-loop filters and on the other hand, displays are equipped with high capacity processing units for post-treatment of decoded videos. This paper proposes a \gls{cnn}-based post-processing algorithm for intra and inter frames of \gls{vvc} coded streams. Depending on the frame type, this method benefits from normative prediction signal by feeding it as an additional input along with reconstructed signal and a \gls{qp}-map to the \gls{cnn}. Moreover, an optional \gls{ms} strategy is adopted to pick the best trained model among available ones at the encoder side, and signal it to the decoder side. This \gls{ms} strategy is applicable at both frame level and block level. The experiments under the \gls{ra} configuration of the \gls{vvc} Test Model (VTM-10.0) show that the proposed prediction-aware algorithm can bring an additional \acrshort{bd-br} gain of -1.3\% compared to the method without the prediction information. Furthermore, the proposed \gls{ms} scheme brings -0.5\% more \acrshort{bd-br} gain on top of the prediction-aware method. 
\end{abstract}



\section{Introduction}
\noindent

Modern block-based video codecs, widely used in nowadays multimedia broadcast industry, often introduce various types of artifacts and distortions. These artifacts are more visible and undesirable in low bitrate video coding, where coarser quantization is applied on transform coefficients in order to drastically reduce the rate. \gls{qe} methods have proved to be efficient for addressing this problem by showing promising performance in compressed videos and have attracted a lot of attention in recent years in the form of \gls{ilf} or \gls{pp}.

Existing video coding standards benefit from different \gls{qe} filters in their compression algorithms as in-loop filter. In \gls{vvc}, the upcoming video coding standard, three in-loop filters are integrated \cite{vvcinloop}, namely \gls{dbf}, \gls{sao} and \gls{alf}. In \gls{dbf}, 1-D low pass filters smooth out block borders and correct the discontinuous edges across them. In \Gls{sao}, reconstructed pixels are categorized into pre-trained classes and are associated with a set of optimized offsets to be transmitted for texture enhancement. \Gls{alf}, that is applied after \gls{dbf} and \gls{sao} in \gls{vvc}, further enhances the quality of the reconstructed frames. In \gls{alf}, parameters of a set of low pass filters are optimized at the encoder side and transmitted to the decoder.

Studies in recent years, mostly focus on the capacity of \glspl{cnn} in different image and video processing problems and exploring new areas of improvement. In the domain of compressed videos \gls{qe}, notable \glspl{cnn}-based works are proposed which provide significant improvements in terms of subjective or objective quality. These methods can be categorized into two general groups: 1) single frame and 2) multiple frame methods. In single frame group\cite{ma2020mfrnet}, video frames are enhanced frame by frame only taking into account the spatial information of each frame. In the second group, multiframe methods\cite{yang2018multi,guan2019mfqe}, the temporal aspect of the video is also taken into account for \gls{qe} task. In addition, some methods~\cite{he2018enhancing,wang2019attention, huang2020frame,nasiri2020prediction} in these two groups, exploit coding information, extracted from bitstream, in order to further improve the quality. Such coding information (\textit{e.g.} \gls{qp}, partitioning, prediction) can help the network to more efficiently remove the coding artifacts. 

In this paper, we proposed a \gls{cnn}-based \gls{qe} method where two coding information of \gls{qp} and prediction signal are used to further improve the performance. The proposed \gls{qe} method is employed as \gls{pp} of \gls{vvc}, both on intra and inter frames. Multiple models for different frame types are trained and a \gls{ms} strategy is employed at the encoder side to choose the best model and signal it in the frame level or \gls{ctb} level for the luma component.

The rest of this paper is organized as follows. In Section~\ref{sec:proposed}, the proposed \gls{qe} method using coding information and the \gls{ms} is presented. Experimental results as well as discussions and comparisons with state-of-the-art solutions are provided in Section~\ref{sec:conclusion} and finally, Section \ref{sec:conclusion} concludes the paper.


\section{Proposed Quality Enhancement method}
\label{sec:proposed}
In this section, first, the coding information exploited in the training and the inference of the proposed \gls{qe} method is explained. Then, the network architecture to integrate this information is presented and finally, a \gls{ms} strategy at the frame level and the \gls{ctb} level is proposed. 

\subsection{Coding information}
Normative decisions made by an encoder are results of extensive searches over possible values of parameters corresponding to its internal coding tools. To make optimal decisions, the spatial and temporal features of video as well as bandwidth constraints are taken into account. Consequently, the information associated to these coding decisions provide rich and informative representation of signal characteristic. Therefore, in the proposed \gls{qe} method, we exploit two main coding information to help our \gls{cnn} networks better remove compression artifacts.


\subsubsection{QP-map}
The quantization step, determined by \gls{qp}, controls the balance between the level of distortion and the bitrate of a compressed video. Higher \gls{qp} values apply coarser quantization step on transform coefficients which results in throwing out more high frequency information, hence less bitrate and higher distortion. 

In our proposed \gls{qe} method, we construct a normalized \gls{qp}-map for each frame and feed it to the network at the same stage as the reconstructed frame. The normalized \gls{qp}-map ($\mathcal{Q}$) for a frame with the width and height of $W$ and $H$, respectively, is calculated as:
\begin{equation}
    \mathcal{Q}_{i,j} = \frac{q_{i,j}}{q_{max}}, 
\end{equation}
where $q_{i,j}$ is the \gls{qp} value of the block that contains the pixel at coordinates $(i,j)$, with $0\leq i < W; 0 \leq j < H$, and $q_{max}$ is the maximum \gls{qp} value (\textit{e.g.} 63 in \gls{vvc}). In constant \gls{qp} mode, as in the \gls{jvet} \gls{ctc} \cite{JVETT2006}, the \gls{qp}-map of a frame would contain a constant value. However, in the \gls{cbr} mode, this value may change at the block level.


\subsubsection{Prediction signal}
The reconstructed signal is roughly the addition of the prediction and the quantized residual signals. Depending on the coding type of the signal, whether intra or inter, different prediction schemes can be used throughout a video: intra, uni-direction inter, bi-direction inter, skip, \textit{etc}. In all these cases, the normative prediction signal determined by the encoder is used as the second coding information in the proposed \gls{qe} method. 

Intra coding is based on spatial redundancies existing within a frame. In \gls{vvc}, a set of 67 \gls{ipm}, representing 65 angular modes, plus \gls{dc} and planar modes are used for modelling texture of blocks. The selection of an \gls{ipm} for a block is performed by optimizing the rate-distortion (R-D) cost. 
Due to the complex R-D optimization process, the generated intra prediction can significantly affect the distortion pattern of the reconstructed block~\cite{nasiri2020prediction}. Thus, bewaring the \gls{qe} networks of the prediction signal can potentially help it learn correlation between the prediction signal and the distortion patterns. 

\begin{figure}
    \centering
    \input{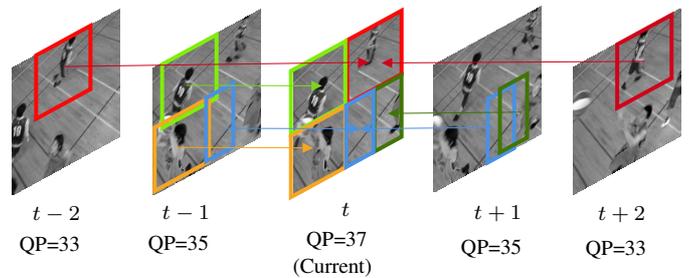}
    \caption{An example of \gls{qp} cascading in the hierarchical GoP structure, providing higher quality motion compensated blocks at frame $t$ from past ($t-1$ and $t-2$) and future ($t+1$ and $t+2$) frames. Each block in frame $t$ is predicted from at least one reference frame with lower \gls{qp} (\textit{i.e.} higher quality texture information).}
    \label{fig:inter}
\end{figure}

Inter coding, on the other hand, is based on temporal redundancies, between consecutive video frames. The prediction signal in the inter mode is a block, similar to the current one, selected from within the range of \glspl{mv} search, based on a distortion metric. In modern video codecs, the search for such similar block is performed inside multiple reference frames that are in the temporal neighborhood of current frame. Fig. \ref{fig:inter} visualizes how the prediction information signal in an inter frame is concatenated from its reference frames. In this figure, current frame at time $t$ uses four reference pictures, two from the past ($t-1$, $t-2$) and two from the future ($t+1$, $t+1$). This usually allows finding very accurate displaced versions of the content in the current block, that can potentially be exploited for the \gls{qe} task.

Regardless of the coding type, the process of generating the prediction signal in the proposed method is the same. To do so, for each reconstructed frame, the prediction blocks of individual \glspl{cu} within it are concatenated. This forms a prediction signal of the same size as the reconstructed frame and is fed to the \gls{qe} network.


\subsection{QE Network}

\begin{figure*}
    \centering
   \scalebox{0.9}{ \input{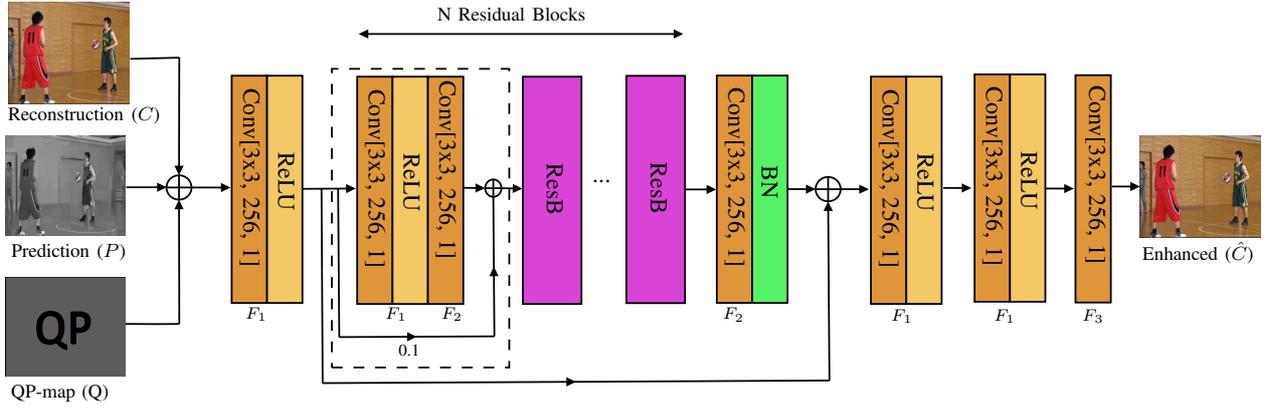}}
    \caption{Network architecture of the proposed method using the prediction, QP-map and reconstruction signal as the input.}
    \label{fig:arch}
\end{figure*}

The network architecture for proposed \gls{qe} method is based on residual blocks which have showed promising performance in low level image processing tasks such as \gls{sr} \cite{lim2017enhanced}. In Fig. \ref{fig:arch} the different layers and skip connections of the used architecture are shown. 

Given $\mathcal{I}$ as the concatenation of the input signals, the process of producing the enhanced reconstructed signal $\mathcal{\hat{C}}$, by the proposed \gls{cnn}-based \gls{qe} method is summarized as: 
\begin{equation}
\label{eq:verbose}
\mathcal{\hat{C}}= F_3^1(F_1^2(Bn^1(F_2^1(Res^{N}(F_1^1(\mathcal{I})))) + F_1^1(\mathcal{I}))),
\end{equation}
where $F_1(.)$ and $F_2(.)$ are $3\times 3 \times 256$ convolutional layers, with and without the ReLU activation layer, respectively. Moreover, $F_3(.)$ is a $3\times 3 \times 1$ convolutional layer with the ReLU activation layer. The superscript of each function indicates the number of times they are repeated sequentially in the network architecture. Finally, $Res$ and $Bn$ are the residual block and batch normalization layer, respectively.

Based on the above network architecture and the use of the prediction information, four models are trained with different sets of inputs. Each model is trained using frames that are encoded in its given coding mode and also used for the inference of the same coding mode. In the first two models, denoted as $M_{intra}^{cqp}$ and $M_{inter}^{cqp}$, the input is the concatenation of the decoded image $\mathcal{C}$, the \gls{qp}-map $\mathcal{Q}$ and the prediction signal $\mathcal{P}$:

\begin{equation}
\label{eq:predaware}
\mathcal{I}_{cqp}^{m}=\mathcal{C}^{m} \oplus \mathcal{Q \oplus P}^{m}, 
\end{equation}
where $\oplus$ is the concatenation operator and \textit{m} is the coding mode which can be \textit{intra} or \textit{inter}. The other two models, denoted as $M_{intra}^{cq}$ and $M_{inter}^{cq}$, do not use the prediction signal as input:    
\begin{equation}
\label{eq:predunaware}
\mathcal{I}_{cq}^{m}=\mathcal{C}^{m} \mathcal{\oplus Q},
\end{equation}

\begin{table}
    \centering
    \caption{Summary of the four models trained for different coding types.}
    \begin{tabular}{lcccc}
        \hline 
         \multirow{2}{*}{Name}      & \multicolumn{3}{c}{Inputs}     & Frames \\
         \cline{2-4}
                                    & Reconstruction & Quantization & Prediction & type \\
         \hline
         \hline
         $M_{cq}^{intra}$           & \checkmark & \checkmark & \ding{55}                   & Intra \\
         $M_{cqp}^{intra}$          & \checkmark & \checkmark & \checkmark                  & Intra \\
         $M_{cq}^{inter}$           & \checkmark & \checkmark & \ding{55}                   & Inter \\
         $M_{cqp}^{inter}$          & \checkmark & \checkmark & \checkmark                  & Inter \\
        \hline
    \end{tabular}
    \label{tab:models}
\end{table}

Table \ref{tab:models} summarizes the details of four proposed models. In all four models, the \gls{qe} task can be formulated as:  

\begin{equation}
\label{eq:summary}
\mathcal{\hat{C}}=f_{QE}(\mathcal{I}; \theta_{QE}),
\end{equation}
where $\theta_{QE}$ is the set of parameters in the network architecture of Eq.~\eqref{eq:verbose}. This parameter set is optimized in the training phase, using the $L_1$ norm as the loss function, computed with respect to the original signal $\mathcal{O}$:

\begin{equation}
    L_1(\mathcal{O}, \mathcal{\hat{C}})=|\mathcal{O}-\mathcal{\hat{C}}|.
\end{equation}


\subsection{Model Selection (MS)}

\begin{figure*}
    \centering
    \input{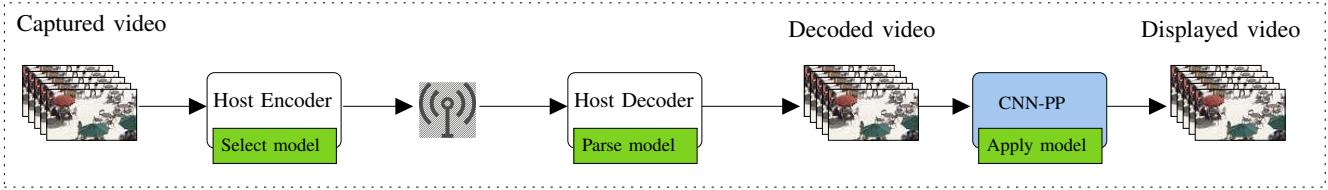}
    \caption{The workflow scheme of the proposed  \gls{cnn}-based \gls{pp} with explicit \gls{ms}}
    \label{fig:model_selection}
\end{figure*}

As discussed, based on frame type, the prediction signal used for blocks of a frame can be different. In intra frames, all blocks are encoded using the intra coding mode. However, in inter frames, blocks can be either inter coded or intra coded, depending on the local motion and texture complexity. Moreover, in all frame types, there are often blocks whose residual signal is zero, which makes the prediction signal identical to the reconstructed block. 
As a result, different types of artifacts can be found in the same encoded frame, that might need different networks for enhancement. Using the four models presented in previous section, a \gls{ms} strategy is proposed in two levels: frame and \gls{ctb}. 

At the \gls{ctb} level, each \gls{ctb} is enhanced by all four models at encoder side. For a given \gls{ctb}, with the original content $\mathcal{O}$, the \gls{ms} at the \gls{ctb} level is performed by minimizing the \gls{mse} as:

\begin{equation}
\label{eq:modelselection}
    M_{in^*}^{m^*} \text{ : }(in^*, m^*) = \operatorname*{argmin}_{m, in} \; \text{MSE}(\mathcal{\hat{C}}_{in}^{m}, \mathcal{O}),
\end{equation}
where $\mathcal{\hat{C}}_{in}^{m}$ is the enhanced signal using the model $M_{in}^{m}$. 

In order to inform the decoder about the best model chosen by encoder, the corresponding information should be transferred in the bitstream. To this end, a signaling scheme is implemented at both \gls{ctb} and frame levels. The frame level signaling is performed with a flag $f_1$ to indicate whether or not the \gls{ctb} level signaling is used. In the case that this flag is zero, decoder will use a default model, either $M_{cqp}^{intra}$ or $M_{cqp}^{inter}$, depending on the frame type. Otherwise, the selected model is determined in the \gls{ctb} level, using two flags $f_2$ and $f_3$. The encoder side decision for the \gls{ms} is presented in Algorithm 1. The inputs to this algorithm are the rate and distortion of the encoded frame, enhanced by the default model, denoted as $R^{def}$ and $D^{def}$, respectively.  Moreover, Fig. \ref{fig:model_selection} demonstrates the overall workflow of proposed \gls{qe} with explicit \gls{ms}.  

\begin{algorithm}
\caption{Frame level of \gls{ms} }
\begin{algorithmic}
\STATE $\textbf{input: } R^{def}, D^{def} \text{ as rate and dist. of frame, respectively.}$
\STATE $R_{f_1:0} \leftarrow R^{def} + 1$
\STATE $R_{f_1:1} \leftarrow R^{def} + 1 + 2 \times (\text{number of CTBs in frame})$
\STATE $D_{f_1:0} \leftarrow D^{def}$
\STATE $D_{f_1:1} \leftarrow 0$

\FOR{ \textbf{each} $\text{CTB } u \in \text{frame}$}
    \STATE $ \text{Get } M_{in^*}^{m^*} \text{ for \textit{u} using Eq.
    \eqref{eq:modelselection}}$
    \STATE $\text{Set } f_2 \text{ and } f_3 \text{ based on } M_{in^*}^{m^*}$
    \STATE $\text{Enhance } u \text{ with } M_{in^*}^{m^*} \text{ using Eq. \eqref{eq:summary}}$
    \STATE $\text{Compute } D_u \text{ as the distortion of \textit{u} after enhancement}$  
    \STATE $D_{f_1:1} \leftarrow D_u $
\ENDFOR
\STATE $J_{f_1:1} \leftarrow D_{f_1:1} + \lambda R_{f_1:1}$
\STATE $J_{f_1:0} \leftarrow D_{f_1:0} + \lambda R_{f_1:0}$
\IF{$J_{f_1:1} < J_{f_1:0}$}
\STATE $f_1 \leftarrow 1 $
\ELSE
\STATE $f_1 \leftarrow 0 $
\ENDIF

\label{alghModelselection}
\end{algorithmic}
\end{algorithm}


\section{Experimental Results}
\label{sec:results}
\subsection{Datasets and training configuration}
For training the networks, the BVI-DVC dataset \cite{ma2020bvi}, recommended by recent Deep Neural Network Video Coding (DNNVC) \glspl{ctc} \cite{JVETT2006, JVETT0129}, has been used, which consists of 800 videos. Moreover, two image datasets, namely DIV2K and Fliker2K are also used for training the networks for intra frames, which are composed of 900 and 2650 high quality images, respectively. The videos and images in the training dataset were converted to 10-bit YCbCr 4:2:0 and only the luma component has been used for training. 

To train the $M_{cq}^{inter}$ and $M_{cqp}^{inter}$ models, the videos were encoded with VTM-10 in \gls{ra} (Main10 profile) in four \gls{qp} values (\{22, 27, 32, 37\}). Moreover, to train $M_{cq}^{intra}$ and $M_{cqp}^{intra}$ models, the DIV2K and Fliker2K datasets were encoded in the \gls{ai} configuration in same four \glspl{qp}. In total, 16000 inter and 7500 intra frames were obtained for training the corresponding models. The training has been performed on $64 \times 64$ patches, randomly chosen from the training dataset. These patches are fed to the network on batches with a size of 16. Block rotation and flip were also applied randomly to the selected patches in order to achieve data augmentation. 

For the test phase, nineteen sequences from the JVET \glspl{ctc} (classes A1, A2, B, C and D ) were used \cite{JVETN1010}. These videos were also encoded in \gls{ra} configuration. Also, none of these videos are included in training phase. 

The networks were implemented in PyTorch platform and the training was performed on NVIDIA GeForce GTX 1080Ti GPU. The parameter $N$ (number of residual blocks of the network) was set to 16. All networks were trained offline before encoding. The initial learning rate was set to $10^{-5}$ with a decay of $0.5$ for every 100 epochs. The Adam optimizer \cite{kingma2014adam} was used for back propagation during the training and each network was trained for 500 epochs. The validation dataset was extracted from the training dataset and was composed of 50 cropped reconstructed frames and their corresponding prediction and original frames. 


The original VTM-10 with all in-loop filters activated is used as the anchor for all test. Moreover, to show the effectiveness of each contribution separately, the performance of three versions of the proposed method are presented: 1) prediction-unaware without \acrfull{ms}, 2) prediction-aware without \gls{ms} and finally 3) prediction-aware with \gls{ms}. To measure the performance, the \gls{bd-br} \cite{bjontegaard2008} metric was used which is formally interpreted as the amount of bit-rate saving at the same level of \gls{psnr}-based quality as the anchor. 

\begin{table}[!b]
\centering
\caption{\acrshort{bd-br} metric for performance comparison of the three versions of the proposed \gls{cnn}-based \gls{qe} method as \gls{pp} in the RA coding configuration of VTM-10}

\renewcommand{\arraystretch}{1.15} 
\begin{tabular}{P{0.02\textwidth}C{0.07\textwidth}P{0.09\textwidth}P{0.08\textwidth}P{0.12\textwidth}}
\hline
\hline

Class & Sequences & Pred-unaware & Pred-aware & Pred-aware + \gls{ms} \\ 
\hline
\multirow{4}{*}{A1} 	&	Tango	        &  -5.92\%	 &	-8.40\%	&	-8.43\%  \\	
                    	&	FoodMarket	    &  -4.46\%	 &	-7.04\%	&	-7.39\%  \\	
                    	&	CampFire	    &  -4.20\%	 &	-5.17\% &	-6.73\%  \\	
                    	&	Average	        &  -4.86\%	 &	-6.87\%	&	-7.52\%  \\	\hline
\multirow{4}{*}{A2} 	&	CatRobot	    &  -6.96\%	 &	-8.41\%	&	-8.58\%  \\	
                    	&	Daylight	    &  -9.33\%	 &  -10.66\%	&	-10.92\% \\	
                    	&	ParkRunning	    &  -3.06\%	 &	-4.04\% &	-4.25\%  \\	
                    	&	Average	        &  -6.45\%	 &	-7.71\% &	-7.92\%  \\	\hline
\multirow{6}{*}{B}  	&	MarketPlace	    &  -4.80\%	 &	-5.54\% &	-5.55\%  \\	
                    	&	RitualDance	    &  -6.15\%	 &	-7.76\% &	-8.06\%  \\	
                    	&	Cactus	        &  -4.21\%	 &	-5.95\% &	-6.62\%  \\	
                    	&	BasketballDrive	&  -5.19\%	 &	-6.73\% &	-7.44\%  \\	
                    	&	BQTerrace	    &  -4.87\%	 &	-5.89\%	&	-7.06\%  \\	
                    	&	Average	        &  -5.04\%	 &	-6.38\% &	-6.95\%  \\	\hline
\multirow{5}{*}{C}  	&	BasketballDrill	&  -6.47\%	 &	-8.27\%	&	-8.56\%  \\	
                    	&	BQMall	        &  -5.13\%	 &	-6.60\% &	-7.31\%  \\	
                    	&	PartyScene	    &  -5.11\%	 &	-5.74\% &	-7.04\%  \\	
                    	&	RaceHorses	    &  -2.71\%	 &	-4.10\% &	-4.24\%  \\	
                    	&	Average	        &  -4.86\%	 &	-6.18\% &	-6.79\%  \\	\hline
\multirow{5}{*}{D}  	&	BasketballPass	&  -7.98\%	 &	-8.79\% &	-8.99\%  \\	
                    	&	BQSquare	    &  -12.39\%	 &	-12.58\%&	-13.10\% \\	
                    	&	BlowingBubble	&  -6.17\%	 &	-6.94\% &	-7.10\%  \\	
                    	&	RaceHorses	    &  -5.59\%	 &	-7.35\% &	-7.35\%  \\	
                    	&	Average	        &  -8.03\%	 &	-8.92\% &	-9.14\%  \\	\hline
				  \multicolumn{2}{c}{All}	&  -5.83\%    &	-7.16\%	&	-7.62\% 
\\ 
\hline
\hline
\end{tabular}
\label{tabbdr2}
\end{table}

\begin{table*}[!t]
\caption{\acrshort{bd-br} performance comparison of the proposed \gls{cnn}-based \gls{qe} methods as \gls{pp} in the RA coding configuration, implemented in VTM-10.}
\label{tab:results}
\centering
\scalebox{0.9}{\renewcommand{\arraystretch}{1.15} 
\begin{tabular}{P{0.03\textwidth}P{0.12\textwidth}P{0.12\textwidth}P{0.12\textwidth}P{0.12\textwidth}P{0.1\textwidth}|P{0.09\textwidth}P{0.12\textwidth}}
\hline
\hline

\multirow{3}{*}{Class}	& \multicolumn{5}{c|}{ State-of-the-art}					 & \multicolumn{2}{c}{Proposed} \\ 
\cline{2-8} 
						 &JVET-O0132 \cite{JVETO0132}        & JVET-O0079 \cite{JVETO0079}   & Zhang \textit{et al.} \cite{zhang2020enhancing}        & JVET-T0079  \cite{JVETT0079} & MFRNet \cite{ma2020mfrnet}   & Pred-unaware      & Pred-aware + MS \\
						& VTM-4			& VTM-5		          & VTM-4				  				& VTM-10				   &VTM-7			                  & VTM-10		              & VTM-10			     		\\

\hline	                  
A1    				   	& -0.15\%                   & -0.87\%          & -2.41\%                            & -2.86\%                      & -6.73\%                  & -4.47\%            & -7.52\%                        \\  						
A2    					& -0.28\%                   & -1.68\%          & -4.22\%                            & -2.98\%                      & -7.17\%                  & -6.41\%            & -7.92\%                        \\  
B     					& -0.22\%                   & -1.47\%          & -2.57\%                            & -2.92\%                      & -6.30\%                  & -5.06\%            & -6.95\%                        \\  
C     					& -0.59\%                   & -3.34\%          & -3.89\%                            & -2.96\%                      & -6.00\%                  & -4.97\%            & -6.79\%                       \\  
D     					& -0.80\%                   & -4.97\%          & -5.80\%                            & -3.48\%                      & -7.60\%                  & -8.04\%            & -9.14\%                        \\  
\hline                                                                                                                                                    
\textbf{All}            & \textbf{-0.41\%}          & \textbf{-2.47\%} & \textbf{-3.76\%}                    & \textbf{-3.04\%}            & \textbf{-6.70\%}          & \textbf{-5.79\%}   & \textbf{-7.62\%}              \\ 
\hline
\hline
\end{tabular}
}
\end{table*}

\subsection{Performance evaluation}
In Table \ref{tabbdr2} the compression performance of the proposed \gls{qe} method in different settings is shown in terms of \gls{bd-br}. As can be seen, the proposed methods in all three settings provide a significant improvement in coding gain on all test sequences with average of -5.83\%, -7.16\% and -7.62\% in pred-unaware, pred-aware and pred-aware + \gls{ms} settings, respectively. It can be observed that adding prediction information brings -1.33\% more bitrate saving to the base \gls{qe} setting. Moreover, using \gls{ms} strategy adds -0.46\% more bitrate saving, where in some sequence like CampFire, BQTerrace and PartyScene the gain is more than -1.3\% in average. 

In order to further investigate the performance of the proposed method, a set of recent \gls{cnn}-based \gls{qe} methods are compared to our proposed method. Two academic papers \cite{ma2020mfrnet,jin2020post} and three JVET contributions \cite{JVETT0079, JVETO0132, JVETO0079} have been selected for the comparison. It should be noted that the coding gains of these works are extracted from their corresponding text. Table \ref{tab:results} summarizes the coding gain of our proposed method as well as aforementioned state-of-the-art methods. First, it can be observed that when our proposed \gls{qe} is integrated as \gls{pp} into the VTM, it outperforms all the competing methods. The performance improvement is consistent over all classes of resolutions. Secondly, the coding gain of our prediction-unaware setting is less than the MFRNet, in which a more efficient network architecture with feature review aspect is deployed\cite{ma2020mfrnet}. It can be concluded that by adding the same strategies to the network of MFRNet, one can obtain even higher performance. This subject is left to be studied in future. 

In terms of complexity, when the proposed \gls{qe} method is used with the \gls{ms} strategy, it adds complexity to the encoder side, unlike most of the \gls{pp} methods. It should be noted that the encoding run-time decreases significantly when a dedicated GPU is available. Finding a strategy to implicitly set the proper model for \gls{qe} task will reduce this complexity at encoder side that is left for a future work.

\section{Conclusion}
\label{sec:conclusion}
\noindent
A \gls{cnn}-based \gls{qe} method for \gls{pp} of \gls{vvc} was proposed. This method is applied on inter and intra coded frames by using prediction information to further improve the performance. In addition, a \acrfull{ms} strategy which selects the best model in terms of R-D cost was employed at the encoder side. The models are trained separately for different frame types (intra, inter) that use particular prediction schemes. The experimental results showed the superior performance of our proposed method compared to the anchor VTM-10 as well as state-of-the-art methods. 

\bibliographystyle{IEEEtran}
\bibliography{mibib} 

\end{document}